\documentclass[twoside]{LCWS11}
\usepackage[latin1]{inputenc}
\usepackage[dvips]{graphicx,epsfig,color}
\usepackage{wrapfig,rotating}
\usepackage{amssymb,amsmath,array}
\usepackage{cite}
\begin{document}
\title{
Calibration Issues for the CALICE 1m$^3$ AHCAL} 
\author{Jaroslav Zalesak$^1$  for the CALICE Collaboration
\thanks{This work is supported by the Ministry of Education, Youth and Sports of the Czech Republic under 
the projects AVO Z3407391, AVO Z10100502, LC527 and INGO LA09042.}
\vspace{.3cm}\\
1- Institute of Physics, Academy of Science of the Czech Republic \\
 Na Slovance 2, 182 21 Prague - Czech Republic
\vspace{.1cm}\\
}

\maketitle

\begin{abstract}

The CALICE collaboration investigates different
technology options for highly granular calorimeters for detectors
at a future electron-positron collider. One of the devices
constructed and tested by this collaboration is a 1m3 prototype
of a  scintillator-steel sampling calorimeter for hadrons
with analogue readout (AHCAL). The light from 7608 small
scintillator cells is detected with silicon photomultipliers. The
AHCAL has been successfully operated during electron and
hadron test-beam measurements at DESY, CERN, and Fermilab
since 2005.
One of main tasks for the successful data taking  is to establish procedures for 
the equalization of cell responses and calibration of such a large number of channels.
  
\end{abstract}


 \section{Introduction}
 
 The CALICE collaboration~\cite{Zalesak:CaliceWeb} is developing 
a hadronic calorimeter (HCAL) with very high granularity 
for future linear colliders (ILC, CLIC). The collaboration built a 1~m$^{3}$ physics prototype in 
2005--6~\cite{Zalesak:Ahcal} and currently CALICE is building an engineering prototype~\cite{Zalesak:Mathias}.

\begin{wrapfigure}{r}{0.5\columnwidth}
\centerline{\includegraphics[width=0.40\columnwidth]{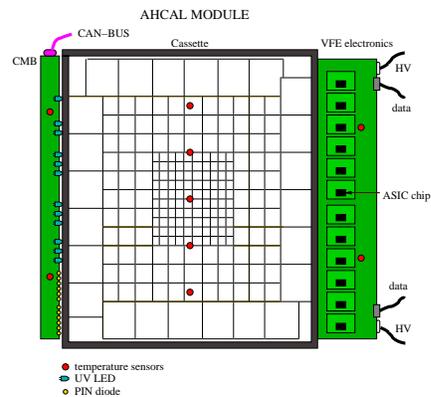}}
\caption{Sketch of one AHCAL module containing the scintillator tiles with SiPM
     readout routed to the VFE electronics (on the right side. The CMB (on the left)
     provides UV LED light for calibration.}
\label{Fig:module}
\end{wrapfigure}

    The AHCAL (Analogue Hadron CALorimeter) 1 m3 physics prototype~\cite{Zalesak:Ahcal} 
with 7608 active readout channels has been in test beams at the CERN SPS in 2006 and 2007, 
followed  in years 2008 and 2009 at Fermilab. The physics prototype is made of 38 layers 
with scintillator tiles, interlaced with 16 mm Fe absorber plates
The dimension of one layer is 90~cm~x~90~cm and it contains scintillator tiles with different granularity 
20~pcs of 12x12~cm$^2$ tiles, 96~pcs of 6x6~cm$^2$ tiles and the first 30~layers have 100~pcs of 
3x3~cm$^2$ tiles in the middle.
The last 8~layers have 6x6~cm$^2$ tiles in the middle instead. 
The active layers are referred to as modules, and the sum of active and passive material 
adds up to a total depth of 5.3 nuclear interaction lengths ($\lambda_i$). 

The active element of the readout is a 
compact photo-detector --- a silicon photomultiplier SiPM
with 1156~pixels, each 32\,$\mu$m$\times$32\,$\mu$m in size. 
The AHCAL analogue board carries 1~ASIC for 18~SiPM channels. A second board
contains control and configuration electronics and provides correct voltage 
for each SiPM. For the calibration a board called CMB is used, see Fig~\ref{Fig:module}.

    The AHCAL readout chain of this prototype contains scintillator tiles with embedded 
wavelength-shifting fibers and small SiPM photo-detectors. The electrical 
signal is adjusted by a preamplifier and a shaper and digitized by a 12-bit ADC. 

  The variation of characteristics of the complete chain (gain, saturation) depends mainly on changes 
of  temperature and operation voltage of the SiPM. For the correct offline reconstruction of the 
energy deposition, the calibration runs have to be included into the data-taking process, since the 
condition inside the detector can change. For purposes of monitoring the long-term stability 
and performance of the photodetectors the calibration  system  with optical fibers for a UV LED light 
distribution was built~\cite{Zalesak:IEEE2010}.
A calibration and monitoring board (CMB) connected to each module distributes UV light from an LED
to each tile via clear fibers. The LEDs are pulsed with 10 ns wide signals steerable in amplitude. 
By varying the voltage, the LED intensity covers the full dynamic range from zero to saturation
(about 70 times the signal of a minimum-ionizing particles --- MIP).
Furthermore, the LED system monitors variations of SiPM gain and signal response, 
both sensitive to temperature and voltage fluctuations. 
The LED light itself is monitored with a PIN photo-diode to correct for fluctuations in the LED
 light intensity.

\section{Calibration procedure}\label{sec:calib}
 
  For trustworthy data measurements with the AHCAL prototype we have to  
establish a reliable and robust calibration chain. This requires measurements
with beam particles and with light from the LED monitoring system. 
The calibration chain is summarized in the following steps:
\begin{itemize}
\item calibration of the cell response and cell-to-cell equalization;
\item monitoring of the SiPM gain and corrections for the non-linear
 response;
\item calibration to an energy scale (in GeV) with electromagnetic showers.
\end{itemize}
For an energy deposit  $E_i$  in units of MIP of one single cell $i$ (with registered signal $A_i$ in ADC counts)
we can write

\begin{equation}
  \label{eq:calib}
  E_i\,[{\rm MIP}] =  \frac{A_i\,[{\rm ADC}]}{C_i^{\rm{MIP}}} \cdot
  f_{\rm{sat}}(A_i\,[{\rm pix}]), 
\end{equation}
where the first term is calibrated signal response in a given cell in case of a linear device.
Due to the non-linearity in the SiPM response (finite number of SiPM pixels), we need to 
add a correction function, which is described by the second term in the Eq.~\ref{eq:calib}.  
This correction is a function of a number of SiPM  pixels, $A_i[{\rm pix}]$, firing for a single cell $i$.
It is related to the ADC value for the cell, $A_i[{\rm ADC}]$, and the corresponding SiPM gain, 
$C_i^{\rm pix} [{\rm ADC}]$:
\begin{equation}
  \label{eq:gain}
  A_i\,[{\rm pix}] = A_i\,[{\rm ADC}]/C_i^{\rm pix}[{\rm ADC}].
\end{equation}

The first term of Eq.~\ref{eq:calib} expresses the equalization of all cell responses done 
by the cell-by-cell calibration, see the section~\ref{ssec:mip}. 
The section~\ref{sec:led} describes how to determine the gains for single cells 
with respect of temperature dependence. 
Methods how to correct for the non-linear SiPM response 
are introduced in the section~\ref{sec:nonlinear}.

\subsection{Cell response equalization with MIPs}\label{ssec:mip}

\begin{wrapfigure}{r}{0.5\columnwidth}
\centerline{\includegraphics[width=0.40\columnwidth]{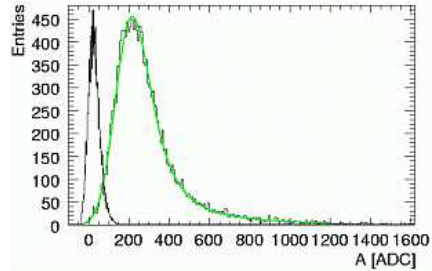}}
\caption{Single calorimeter cell response to muons with the corresponding fit 
(green line) and noise spectrum of the same cell.}
\label{Fig:mipfit}
\end{wrapfigure}

The cell-by-cell calibration (equalization of all cell responses) is performed using 
minimum-ionizing particles (MIPs) provided by a broad muon beam, illuminating all cells 
in the detector. For each cell, a calibration factor, $C_i^{\rm{MIP}}$, is determined 
from the most probable value of the measured energy spectrum for muons in ADC units, 
which is extracted with a fit using a Landau function convoluted with a Gaussian, 
as can be seen for one cell in Fig~\ref{Fig:mipfit}.

This fit accounts for the distribution of energy loss of muons in the scintillator tiles 
as well as for contributions from photon counting statistics and electronic noise. 
The combined systematic and statistical uncertainty for these fits was typically 
on the order of 2\%. The muons are generally parallel to the beam line and perpendicular 
to the detector front face. In this way all cells can be calibrated at the same time, 
minimizing the impact of temperature induced variations. 

The MIP calibration efficiency, i. e. the amount of successful MIP calibrations 
above the 0.5$\times$MIP threshold, is about 93\%
for the 2007 data set (CERN tests). The calibration fails for cells with broken
electrical connection or very noisy SiPMs. 
The signal over noise ratio ($N/S$), defined as MIP
amplitude over pedestal width (see Fig.~\ref{Fig:mipfit}), 
is a measure for the separation of a MIP signal 
and noise. It is found to be $\sim$10 in 2007.
 
 \section{UV LED calibration and monitoring system}\label{sec:led}

 \begin{wrapfigure}{r}{0.5\columnwidth}
\centerline{\includegraphics[width=0.40\columnwidth]{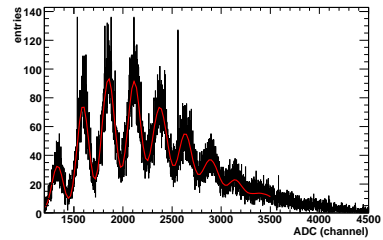}}
\caption{Single photo-electron peak spectrum taken with a SiPM, 
fitted with a multi-gaussian formula (red curve).}
\label{Fig:gainfit}
\end{wrapfigure}

   Since the SiPMs are sensitive to changes in temperature and operation voltage, the LED system has been installed 
to monitor the stability of the readout chain in time. The calibration system needs sufficient flexibility to perform 
several different tasks. Gain calibration: we utilize the self-calibration properties of the SiPMs to achieve the calibration 
of an ADC in terms of pixels that is needed for non-linearity corrections and for a direct monitoring of the SiPM gain. 
Further we monitor all SiPMs during test beam operations with a fixed intensity
light pulse. Saturation: we cross check the full SiPM response function by varying the light intensity from zero to 
the saturation level.
      Our calibration boards use an UV-LED as a light source for calibration. The UV-LEDs require a special 
driver~\cite{Zalesak:Ivo} in order to make them shine fast ($\leq$10~ns) with an amplitude covering several orders of magnitude 
in light intensity.

 \subsection{SiPM gain calibration}\label{ssec:gain}
 
\begin{figure}
\begin{center}
{\includegraphics[width=0.4\columnwidth]{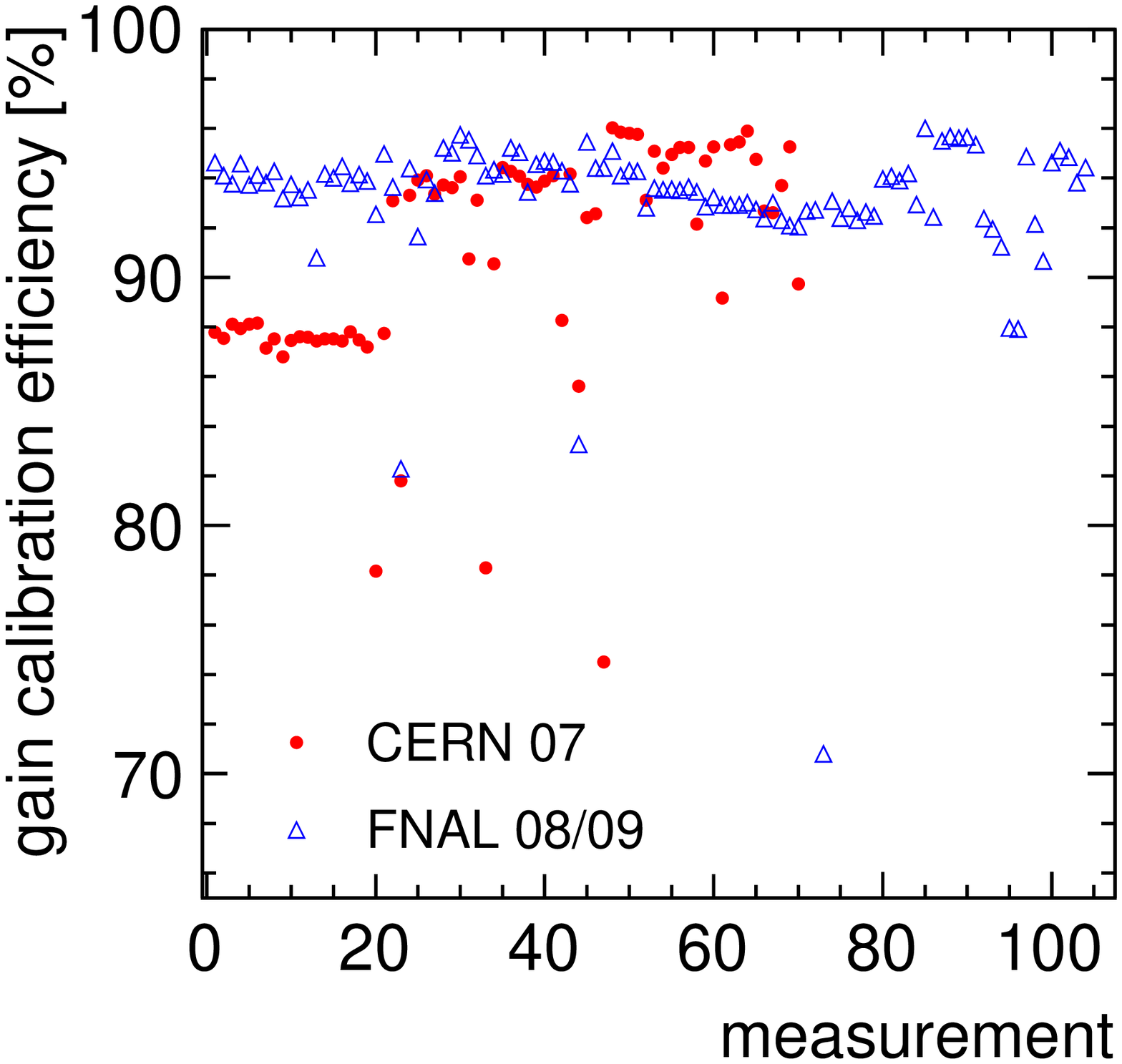}}
{\includegraphics[width=0.4\columnwidth]{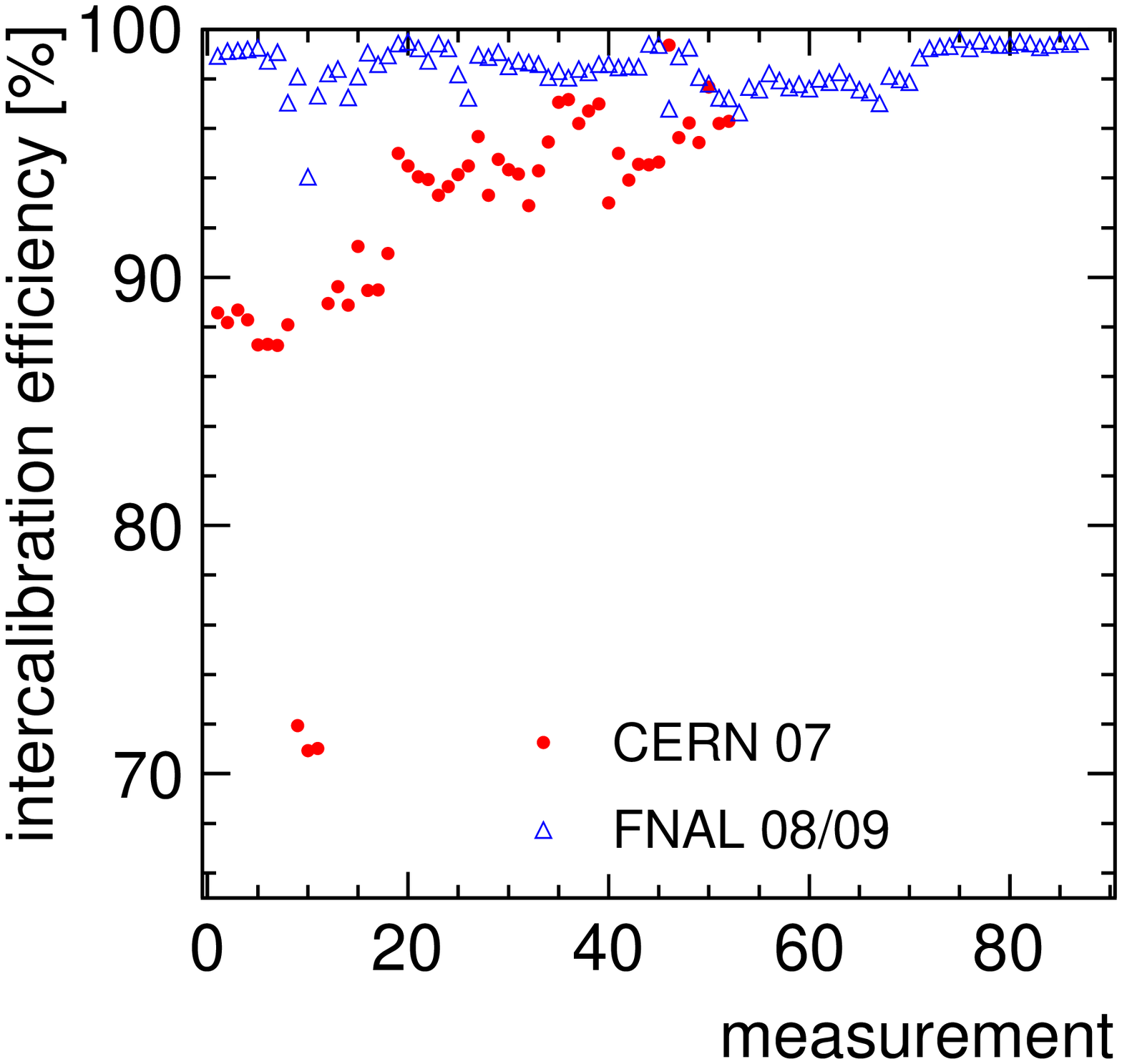}}
\end{center}
\caption{Gain calibration efficiency (left) and  electronics inter-calibration efficiency (right),
      data taken at CERN in 2007  and at FNAL in 2008.
      More than 85.0\,\% of the channels could be
      monitored for gain and inter-calibration variation during these
      periods. }
\label{Fig:EffGainIC}
\end{figure}

For the extraction of the saturation curve~$f_{\rm{sat}}$ (Eq.~\ref{eq:calib}) describing the 
SiPM non-linearity we need to know the SiPM gain for each cell (Eq.~\ref{eq:gain}).
The gain is extracted from single photo-electron (p.~e.)
spectra taken in dedicated runs with low LED light intensity. 
LED light is necessary as the best determination of the gain requires 
a single photo-electron spectrum with a Poisson mean of about 1.5 p.e. 
and the mean obtained from dark noise events is below 0.5 p.e.

The SiPM gain, $G_i^{\rm{fit}}$,
is the distance between two
consecutive peaks in the single photo-electron spectrum. 
A typical gain spectrum is shown in Fig.~\ref{Fig:gainfit}. A multi-Gaussian fit is performed 
to the single photo-electron peaks to determine their average relative distance. 
The mean of each Gaussian function in the multi-Gaussian sum is left as a free parameter and 
the width is dominated by electronic noise, but for large number 
of pixels fired  the statistical contribution becomes visible, which lead to an increase of the peak
width. Accordingly, the width of each peak is left as a free parameter.
The uncertainty on the gain determination is mainly due to the fit and
is about 2\,\% for fits which pass quality criteria.

SiPM gain measurements were repeated approximately every eight hours during 
test beam operation. The SiPM gain varies with temperature and has to be corrected for it.
The efficiency of the gain extraction is defined as the number of successful 
fits in one gain run divided by the number of channels which can be calibrated. 
Figure~\ref{Fig:EffGainIC} (left) shows the efficiency of the gain extraction for a
series of runs taken at CERN and FNAL, respectively. Initial problems during the system 
commissioning phase led to low efficiency, but after commissioning a gain extraction efficiency 
of about~95\,\% per run has been achieved.  The gain efficiency was also stable 
after transportation and throughout the FNAL runs.
Combination of several gain runs yields calibration of more than 99\,\% of all cells. 
The remaining 1\,\% of cells are calibrated with the average of the module to which they belong.

\subsection{ASIC mode inter-calibration}\label{ssec:ic}

The measurement of SiPM gain is performed with a special mode of the readout chip
(high pre-amplification gain, CM mode). 
In contrast, physics data taking are performed with
approximately ten times smaller electronic amplification (PM mode). 
The inter-calibration factor, $I_i$, of the chip gain between both modes
is used to determine the overall SiPM calibration factor
(used in Eq.~\ref{eq:gain}), $C_i^{\rm{pix}}=  G_i^{\rm{fit}}/I_i$.

The extraction of the inter-calibration coefficients depends on 
the linear response of the chip in both modes for an overlapping range of input signals. 
The input signal is provided by the LED system. The amplitude
of the signal is varied within the linear range by varying the LED
light intensity. The response in each readout mode is fit with a line, and the 
ratio between the two slopes is the inter-calibration coefficient for one given
readout channel.
Ideally, this factor should be a simple constant between the two chip readout modes, 
but it turns out to depend on the SiPM signal form due to the different shaping times in the two modes. 
For longer SiPM signals  the inter-calibration is bigger than for shorter SiPM signals. 
The inter-calibration factors between the chip readout modes 
range between 4 and 13.

As with the gain, the inter-calibration extraction efficiency is
influenced by the quality of the LED light distribution system. The
inter-calibration coefficient extraction efficiency during the 2007
and the 2008 data taking periods are plotted in Figure~\ref{Fig:EffGainIC} (right). 
After commissioning was completed, all channels with the exception of the 
2\,\% inactive channels and the channels connected to a broken LED, 
could be inter-calibrated. 
For the missing inter-calibration values the average of the module to which a SiPM belongs is used instead.

 \subsection{Temperature variation}\label{ssec:temp}
 
 The SiPM gain and photo-detection efficiency are temperature dependent. 
The product of the two determines the SiPM response, which typically decreases by 3.7\%/K. 
A procedure has been developed to correct temperature-induced variations in the calorimeter 
response using temperature measurements in each module. This procedure and its stability will 
be described in more detail  in~\cite{Zalesak:muon_paper}. 
To  account for the included temperature variations, the visible energy of each data set 
is scaled by -3.7\%/K to the average temperature of the muon data used for calibration.

 \section{SiPM non-linearity}\label{sec:nonlinear}
 
\begin{figure}
\begin{center}
{\includegraphics[width=0.44\columnwidth]{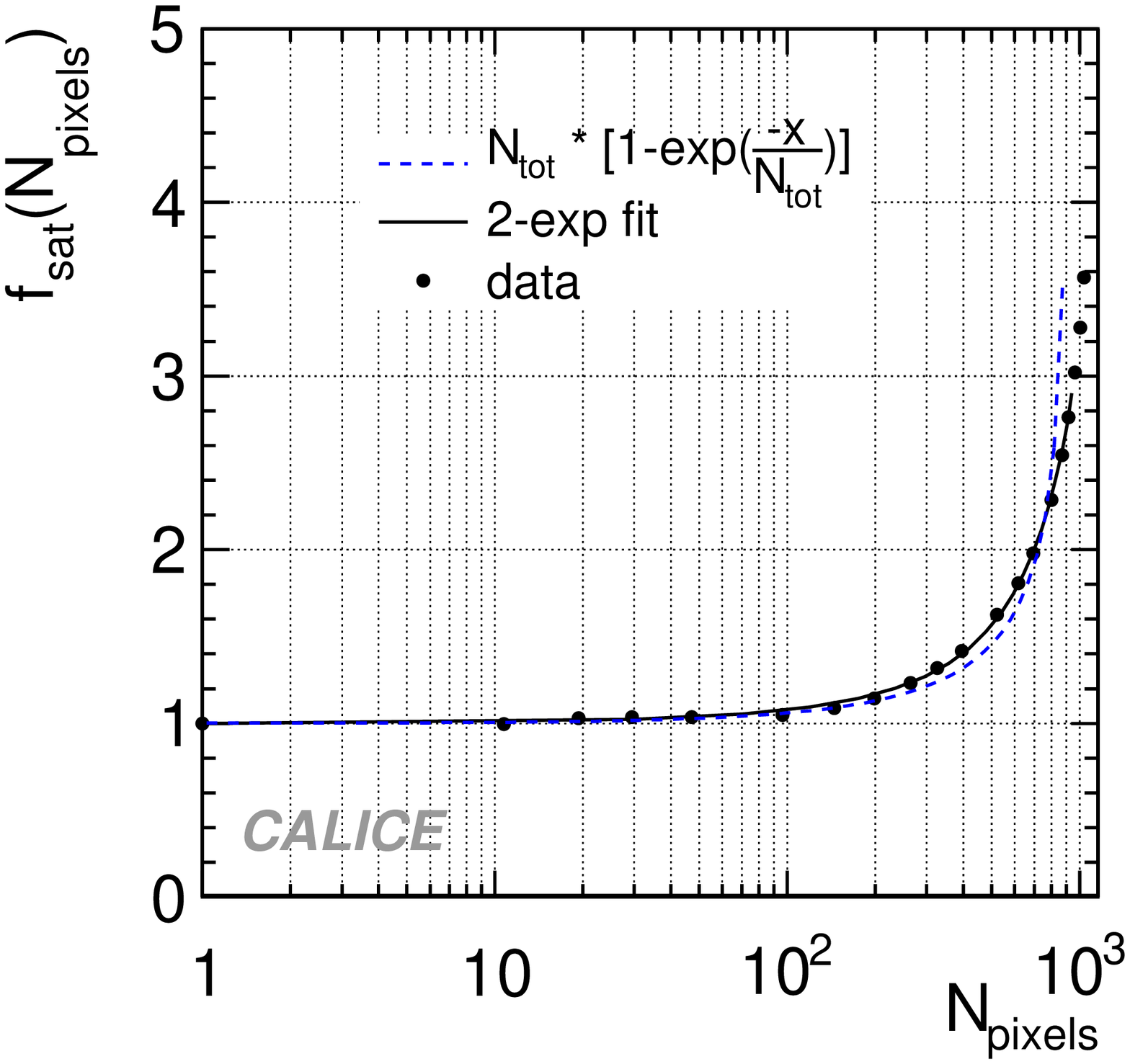}}
{\includegraphics[width=0.44\columnwidth]{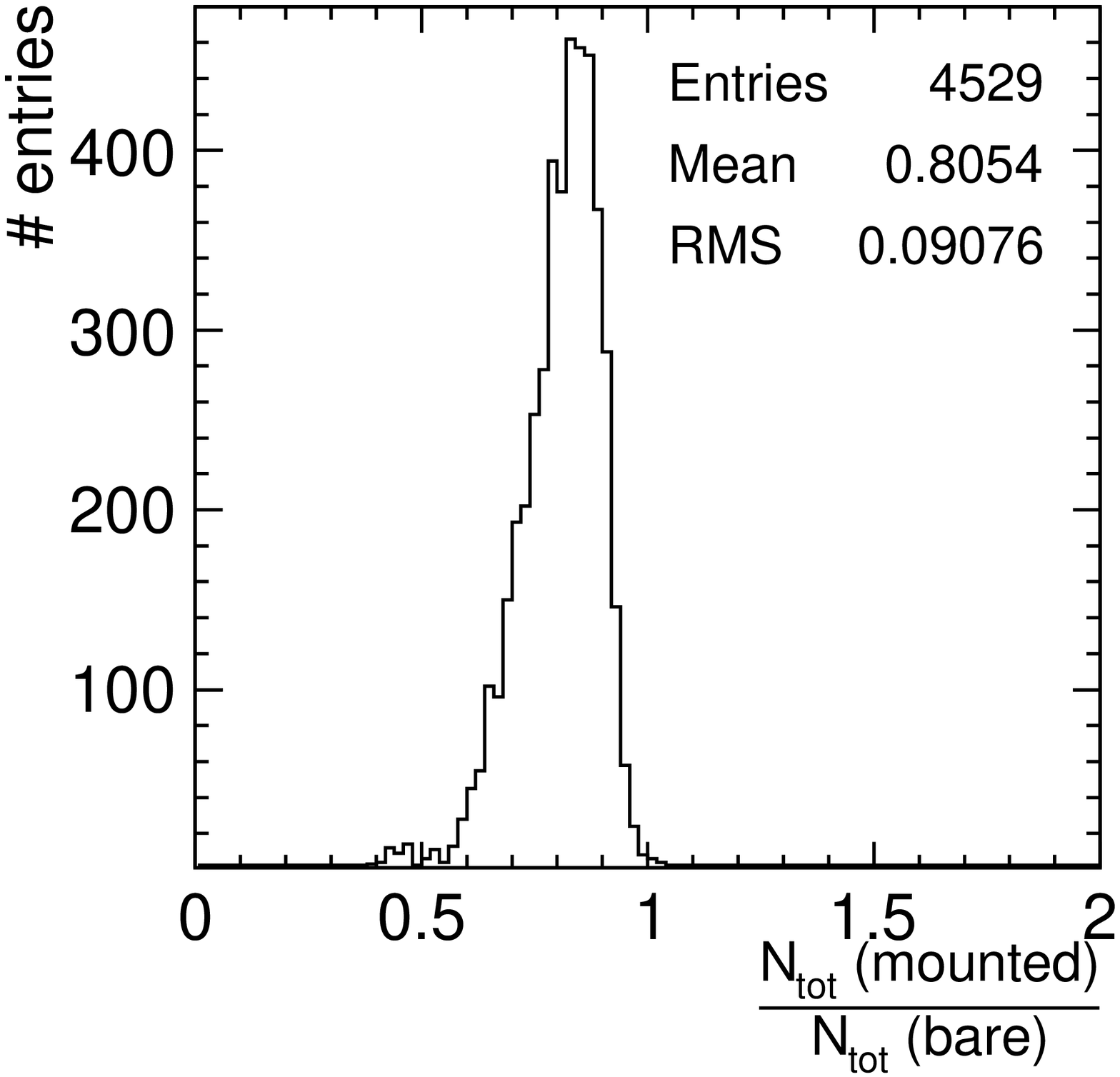}}
\end{center}
\caption{(Left) the SiPM non-linearity correction function, $f_{\rm{sat}}$. The points are the 
                tabulated data of $N_{\rm pix}$ versus $N_{\rm pe}$ for one SiPM in the AHCAL. 
             (Right) Ratio of maximum number of fired pixels, $N_{\rm tot}$(mounted),
               measured with SiPM mounted on a tile to $N_{\rm tot}$(bare) measured 
               directly with bare SiPMs.}
\label{Fig:saturation}
\end{figure}

Due to the limited number of pixels and the finite pixel recovery
time, the SiPM is an intrinsically non-linear device. 
The response function of a SiPM correlates the observed number of pixels 
fired, $N_{\rm pix}$,  to the effective number of photo-electrons generated, 
$N_{\rm pe}$, including cross-talk and after-pulses.
The response of a SiPM can be approximated by the function 
 \begin{equation}
  \label{eq:sat}
  N_{\rm pix} = N_{\rm tot}\cdot(1-e^{-N_{\rm pe}/N_{\rm tot}}),
\end{equation}
with $N_{\rm tot}$ the maximum number of fired pixels.
This formula is a useful approximation for the case of uniform light
distribution over the pixels and short light pulses.  
In the above approximation, one can extract a correction function~$f_{\rm{sat}}(N_{\rm pix})$ 
for the SiPM non-linear response 
as the residual to linearity of the inverted SiPM response function.
For the analysis of measured data we use the tabulated values of a fit of the double exponential function,
 see Fig.~\ref{Fig:saturation}~(left).

\begin{wrapfigure}{r}{0.5\columnwidth}
\centerline{\includegraphics[width=0.45\columnwidth]{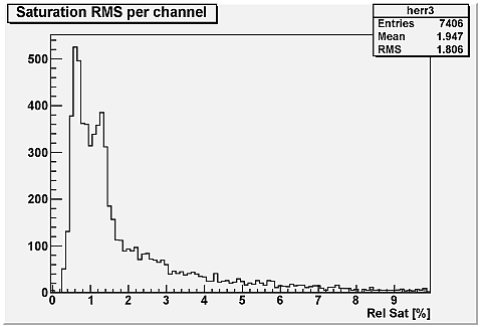}}
\caption{Relative uncertainty of the determination of the saturation point for a single 
channel ($RMS/N_{\rm tot}$). Temperature corrected data covers one full beam test period.}
\label{Fig:sateff}
\end{wrapfigure}

For studies saturation behaviour of the individual SiPMs, they were not mounted on tiles, but were bare SiPMs. Therefore, all
the pixels have been illuminated with light in a homogeneous way. The
 measurement results for all SiPMs installed in the AHCAL are
given in~\cite{Zalesak:Ahcal}. The maximum number of fired
pixels ($N_{\rm tot}$(bare)) for each SiPM is extracted with a fit to the measured points using Eq.~\ref{eq:sat}.
The spread (RMS) in the values of $N_{\rm tot}(bare)$ between all the curves is about 20\,\%. SiPMs with 
$N_{\rm tot}(bare)>$~900 have been pre-selected.
This ensures not too large variations in the non-linear response
function of each device. 
\begin{wrapfigure}{r}{0.5\columnwidth}
\centerline{\includegraphics[width=0.4\columnwidth]{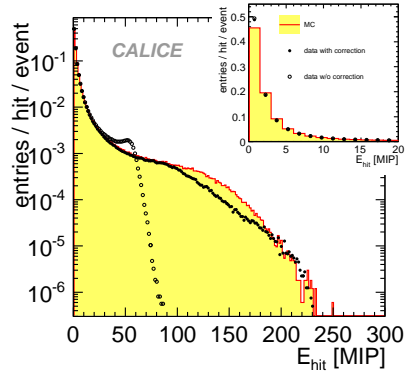}}
\caption{ Hit energy spectrum for 30\,GeV positron showers in the
      AHCAL. Open circles (dots) show the data before (after) 
      correction for the non-linear response of the SiPM.}
\label{Fig:EhitSat}
\end{wrapfigure}

Alternatively, $N_{\rm{tot}}$ has also been extracted using the AHCAL LED monitoring system 
from measurements with the SiPM mounted on a tile ($N_{\rm tot}$(mounted)). 
The saturation point in number of pixels is independent on the linearity of the light.
Fig.~\ref{Fig:saturation} (right) shows the ratio of $N_{\rm tot}$(mounted) to $N_{\rm tot}$(bare).  
The plot shows that the maximum number of pixels in the in-situ setup is on average 80.5\,\% of the
value determined in the laboratory setup~\cite{Zalesak:Nils} with a
wide distribution (RMS=9\,\%). This factor is interpreted as geometric mismatch between
the WLS fiber and the photodetector.  
The fiber has a  1\,mm diameter while the SiPM active surface area is 1$\times$1\,mm$^2$; 
the geometric ratio between areas is 79\,\%, in agreement with the measured value. 
Therefore, only a fraction of the SiPM surface is illuminated and the 
laboratory curves are re-scaled by the measured value of 80.5\,\% to
correct for this effect before they are used to correct for the SiPM
saturation.

The uncertainty of the determination of the saturation point  in-situ ($N_{\rm tot}$)
for a single channel is lower than 3\,\%, if the LED light range properly covers the 
SiPM saturation region, and if this region is measured well below the 
ADC saturation. Unfortunately, these conditions are true only for a sub-sample 
of about 73\,\% channels as can be seen in Fig.~\ref{Fig:sateff}.

The impact of the saturation correction is  seen in Fig.~\ref{Fig:EhitSat}
where the energy per hit is shown with and without the correction factor $f_{\rm{sat}}$ applied, 
for 30\,GeV electromagnetic showers. Whereas the correction is negligible for low signal amplitudes, 
it becomes significant at larger amplitudes, resulting in a strong correction for the tail of the
distribution. For the maximum energy deposited in one cell for a 30\,GeV  the correction factor is
$f_{\rm{sat}}(A_i) \sim 3$.

\begin{wrapfigure}{r}{0.5\columnwidth}
\centerline{\includegraphics[width=0.45\columnwidth]{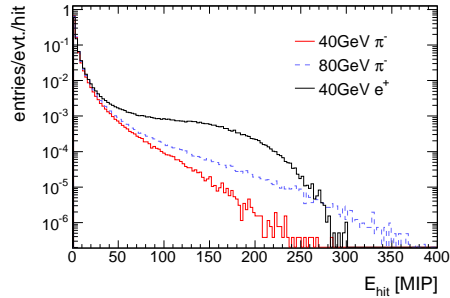}}
\caption{Hit energy spectrum for 40\,GeV positron showers 
       compared to that of 40\,GeV  and 80\,GeV pion showers. }
\label{Fig:Ehit}
\end{wrapfigure}

\section{Validation of the AHCAL calibration}
  
The linearity of the calorimeter response for a large range of incident particle energies is a key feature, 
which allows for an important test of the calibration chain.
Electromagnetic showers offer the most rigorous test for non-linearity correction, 
since the energy deposited per single tile in an electromagnetic
shower is larger than that in a hadronic shower for the same particle energy.
Figure~\ref{Fig:Ehit} shows the hit energy spectrum of a 40\,GeV
positron shower compared to the spectra of 40\,GeV and 80\,GeV pion showers. The positron shower 
clearly has more hits with high
energy deposition, even when the total particle energy is only half that of the pion.

More results about the analysis of the electromagnetic showers measured by the AHCAL
can be found in~\cite{Zalesak:Ahcal2}.

%
%

\section*{Acknowledgments}

The author gratefully thanks to his colleagues from the Institute of Physics in Prague: 
Jaroslav Cvach, Milan Janata, Jiri Kvasnicka, Denis Lednicky, Ivo Polak, Jan Smolik 
and colleagues from the CALICE collaboration for their help and realization the measurements
 needed for this work.
Special thanks go to  Erika Garutti (DESY) for very valuable contributions to my presentation 
at LCWS11 in Granada.


\begin{footnotesize}


\end{footnotesize}


\end{document}